\documentclass[useAMS,usenatbib]{mn2e}

\usepackage{graphicx}

\newcommand{\kepler}{{\it Kepler}}

\title[rms-flux and fast variability of MV\,Lyr]{Rms-flux relation and fast optical variability simulations of the nova-like system MV\,Lyr.}
\author[A. Dobrotka, S. Mineshige and J.-U. Ness]{
A. Dobrotka$^1$\thanks{E-mail: andrej.dobrotka@stuba.sk},
S. Mineshige$^2$\thanks{E-mail: shm@kusastro.kyoto-u.ac.jp}
and J.-U. Ness$^3$\thanks{E-mail: juness@sciops.esa.int}\\
$^1$Slovak University of Technology in Bratislava, Faculty of Materials Science and Technology in Trnava, Institute of Materials\\
Science, Paul\'inska 16, 91724 Trnava, Slovak Republic\\
$^2$Department of Astronomy, Graduate School of Science, Kyoto University, Sakyo-ku, Kyoto 606-8502, Japan\\
$^3$XMM-Newton Science Operations Center, European Space Astronomy Center, PO Box 78, 28691 Villanueva de la Ca\~nada, Madrid, Spain\\
}

\begin{document}

\date{Accepted ???. Received ???; in original form \today}

\pagerange{\pageref{firstpage}--\pageref{lastpage}} \pubyear{2014}

\maketitle

\label{firstpage}

\begin{abstract}
The stochastic variability (flickering) of the nova-like system (subclass of cataclysmic variable) MV\,Lyr yields a complicated power density spectrum with four break frequencies. \citet{scaringi2012b} analysed high-cadence \kepler~data of MV Lyr, taken almost continuously over 600 days, giving the unique opportunity to study multicomponent Power Density Spectra (PDS) over a wide frequency range. We modelled this variability with our statistical model based on disc angular momentum transport via discrete turbulent bodies with an exponential distribution of the dimension scale. Two different models were used, a full disc (developed from the white dwarf to the outer radius of $\sim 10^{10}$\,cm) and a radially thin disc (a ring at a distance of $\sim 10^{10}$\,cm from the white dwarf) that imitates an outer disc rim. We succeed in explaining the two lowest observed break frequencies assuming typical values for a disc radius of 0.5 and 0.9 times the primary Roche lobe and an $\alpha$ parameter of 0.1 -- 0.4. The highest observed break frequency was also modelled, but with a rather small accretion disc with radius of 0.3 times the primary Roche lobe and a high $\alpha$ value of 0.9 consistent with previous findings by \citet{scaringi2014}. Furthermore, the simulated light curves exhibit the typical linear rms-flux proportionality linear relation and the typical log-normal flux distribution. As the turbulent process is generating fluctuations in mass accretion that propagate through the disc, this confirms the general knowledge that the typical rms-flux relation is mainly generated by these fluctuations. In general a higher rms is generated by a larger amount of superposed flares which is compatible with a higher mass accretion rate expressed by a larger flux.
\end{abstract}


\section{Introduction}
\label{introduction}

A basic characteristic of accretion is fast stochastic variability (sometimes also called flickering). Such oscillations are observed in optical and X-rays light curves of several systems as active galactic nuclei, X-ray binaries and cataclysmic variables. This variability reflects as red noise in power density spectra (PDS). The shape of such PDS can be a simple power law with no other components (see e.g. \citealt{shahbaz2010}, \citealt{mushotzky2011}), a broken power law (two power law components) with a single break frequency (see e.g. \citealt{kato2002}, \citealt{baptista2008}) or a multicomponent PDS with more characteristic break frequencies (see e.g. \citealt{sunyaev2000}, \citealt{scaringi2012b}, \citealt{dobrotka2014}). The high-cadence optical \kepler~data of the nova-like system MV\,Lyr, taken almost continuously over 600 days, is a unique opportunity to study such multicomponent PDS over a wide frequency range (\citealt{scaringi2012b}). The PDS has 4 different break frequencies ranging from 0.17 to $13.97 \times 10^{-4}$\,Hz.

A more fundamental characteristic of fast stochastic variability than the power spectrum is the correlation between variability amplitude and log-normally distributed flux (so called rms-flux relation). This relation is valid over a wide-range of timescales. \citet{uttley2005} studied non-linear X-ray variability of X-ray binaries and active galaxies and found a linear relation between rms and flux calculated from light curve segments. \kepler~optical data show the same linear relation in the case of the cataclysmic variable MV\,Lyr (\citealt{scaringi2012a}). Following both publications, the most promising model describing this relation are variations in the accretion rate that are produced at different disc radii (\citealt{lyubarskii1997}, \citealt{kotov2001}, \citealt{arevalo2006}). Therefore, a successful flickering model should also reproduce this rms-flux relation.

Various models were developed in order to simulate flickering activity. A cellular-automaton model was proposed by \citet{yonehara1997} to describe the PDS of flickering in cataclysmic variables. In this model light fluctuations are produced by occasional flare-like events and a subsequent avalanche flow in the accretion disc atmosphere. Flares are assumed to arise when the mass density exceeds a critical limit. This model reproduced the typical broken-power law PDS with a single break frequency (see \citealt{mineshige1994} for the original idea). \citet{dobrotka2010} developed a statistical model to simulate flickering light curves based on the simple idea of angular momentum transport between two adjacent concentric rings in the accretion disc via discrete turbulent bodies. Application to optical observations of the symbiotic system T\,CrB yielded a successful description of the PDS with inner disc truncation, in agreement with \citet{skopal2005}. The derived basic $\alpha$ disc parameter relating kinematic viscosity with the sound velocity and disc scale height (\citealt{shakura1973}) is rather high, but still in the expected interval 0.1 - 1.0, which is typical for hot ionised discs (see e.g. \citealt{kato1998} chapter 5, \citealt{lasota2001}). Further application of the same statistical model to optical data of the nova-like systems KR\,Aur and UU\,Aqr successfully reproduced the PDS of KR\,Aur but not the PDS of UU\,Aqr, perhaps due to spiral arms in the accretion disc of the latter (\citealt{dobrotka2012}). A similar method based on unstable mass flow from the inner disc toward the white dwarf surface modelled fast X-ray stochastic variability of the dwarf nova RU\,Peg in quiescence (\citealt{dobrotka2014}). The observed PDS has a multicomponent shape with two characteristic break frequencies. The model explained the low frequency break as due to viscous processes at the inner disc radius while the high break frequency remains unresolved. Recently, \citet{ingram2013} have derived an analytical expression for the fluctuating accretion rate in the disc, modelling the observed PDS broken power law shape. \citet{scaringi2014} applied the expression to fit the PDS of the nova-like system MV\,Lyr data. The model described the observed PDS with rather non-standard accretion disc parameters, i.e. disc scale height vs. distance from the centre ratio, $H/r>0.1$, while the basic accepted geometrically thin disc model requires $H/r < 0.01$. The authors interpret such disc as expanded optically thin hot corona surrounding an optically thick geometrically thin standard disc.

While \citet{dobrotka2010} and \citet{dobrotka2012} studied only the PDS shape of fast oscillations, the model was not yet used to investigate the physically more important rms-flux relation. In this paper we expand this model to account also for variations in the mass accretion rate through the accretion disc and we search for the multicomponent shape of the PDS in MV\,Lyr together with the rms-flux relation. Section~\ref{mvlyr} describes the knowledge about the MV\,Lyr system and Section~\ref{model} describes the model used for the generation of synthetic light curve. The details of all simulations are described in Section~\ref{simulations} and the results summarised and discussed in Sections~\ref{results} and \ref{discussion}, respectively.

\section{Nova like system MV\,Lyr}
\label{mvlyr}

MV\,Lyr is a nova-like system, a subclass of cataclysmic variables. The accretion discs of these systems are believed to be hot with ionised hydrogen, optically thick, geometrically thin with an $\alpha$ parameter (\citealt{shakura1973}) of the order of 0.1 (see e.g. \citealt{lasota2001} for review). The disc should be developed down to the white dwarf with a very small boundary layer between the inner disc and the white dwarf surface (\citealt{narayan1993}). Therefore, the accretion disc is expected to be at most slightly truncated. This is the opposite of dwarf novae ($\alpha \sim 0.01$) with non-ionised quiescent discs, where the inner disc truncation is expected to be relatively large with a boundary layer that is much thicker in radial direction. The mass accretion rate of MV\,Lyr was probably not constant due to a variable long term trend of the light curve, but \citet{scaringi2012b} assumed an approximate value of $1.9 \times 10^{17}$\,g\,s$^{-1}$ taken from \citet{linnell2005}. The white dwarf has a mass of 0.73\,M$_{\rm \odot}$ (\citealt{hoard2004}) and the orbital period is approximately 3.3\,h (\citealt{borisov1992}).

Typical disc radii in cataclysmic variables are of the order of $10^{10}$\,cm (see e.g. \citealt{warner1995}). The more exact estimate in accreting systems is defined by the tidal radius at 0.9 times the primary Roche lobe radius $r_{\rm PRL}$ (see e.g. \citealt{frank1992}). Taking into account observations of related objects as X-ray binary A0620-00, the disc radii can be estimated as $0.5 \times r_{\rm PRL}$ (\citealt{marsh1994}). The value of $r_{\rm PRL}$ is given by \citet{paczynski1971};
\begin{equation}
r_{\rm PRL} = 0.462~a~\left( 1 + q \right)^{-1/3}~[{\rm cm}],
\end{equation}
where $q$ is the mass ratio and $a$ is the orbital separation calculated from the third Kepler law
\begin{equation}
a = 3.5 \times 10^{10}~\left( \frac{M_1}{\rm M_{\odot}} \right)^{1/3} (1 + q)^{1/3} P_{\rm h}^{2/3}~[{\rm cm}].
\end{equation}
$M_1$ is the primary mass and $P_{\rm h}$ is the orbital period in hours. Combining these two equations, the resulting disc radius is independent of $q$ and thus the secondary mass. Using all other known parameters, the disc radii of 0.5 and $0.9 \times r_{\rm PRL}$ yields 1.8 and $3.2 \times 10^{10}$\,cm, respectively.

\begin{figure}
\includegraphics[width=36mm,angle=-90]{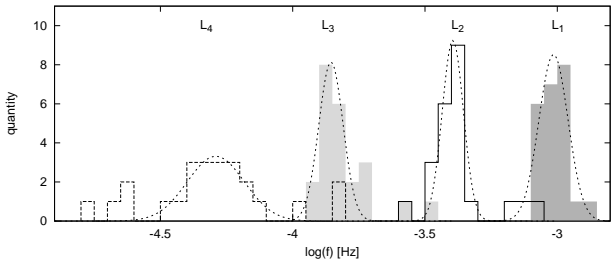}
\caption{Histograms of four break frequencies $L_1$ to $L_4$ derived from \kepler~observations. The dotted lines are Gaussian fits in log-space to the histograms.}
\label{obs_hist}
\end{figure}

The PDS of MV\,Lyr is well fitted with four different Lorentzian functions characterising four different break frequencies $L_1$ to $L_4$ \citep{scaringi2012b}. The authors produced 118 rms-normalised PDSs, each from 5.275-day, non-overlapping data segments. Subsequently, they averaged 5 consecutive PDSs and binned into equally spaced frequency intervals. This resulted in 24 time-averaged PDSs. Every averaged PDS was fitted with four Lorentzian functions and the corresponding histograms of the characteristic frequencies (from Table~1 in \citealt{scaringi2012b}) are depicted in Fig.~\ref{obs_hist}. Clearly the histograms are slightly overlapping. Few values of $L_4$ are extending to the $L_3$ histogram, from $L_3$ to $L_2$ etc. Despite of few data points, we fitted the histograms with Gaussian functions in log space, and the resulting fits are depicted as dotted lines in Fig.~\ref{obs_hist}. Clearly this solves the problem of overlapping histograms. The few marginal data are statistically not significant enough in the fitting procedure and the Gaussian functions describe only the "central" part of each histogram. With this procedure we obtain four well-separated distributions and the mean values with 1-$\sigma$ errors are summarised in Table.~\ref{obs_param}.
\begin{table}
\caption{Mean values of the observed break frequencies $L_{\rm i}$ from Fig.~\ref{obs_hist} with standard deviation of the sample.}
\begin{center}
\begin{tabular}{lccr}
\hline
\hline
$L_{\rm i}$ & log ($L_{\rm i}$/Hz) & $L_{\rm i}$ & log ($L_{\rm i}$/Hz)\\
\hline
$L_1$ & $-3.01 \pm 0.06$ & $L_3$ & $-3.86 \pm 0.05$\\
$L_2$ & $-3.39 \pm 0.04$ & $L_4$ & $-4.29 \pm 0.11$\\
\hline
\end{tabular}
\end{center}
\label{obs_param}
\end{table}

\section{Model}
\label{model}

A detailed description of our numerical modelling can be found in \citet{dobrotka2010}. The method is based on the simple idea of discrete angular momentum transport between two adjacent concentric rings in a geometrically thin disc. The angular momentum is carried by discrete turbulent bodies with an exponential distribution of the dimension scales. When a turbulent body penetrates from one ring to the adjacent one, it changes the tangential velocity and distance from the centre, while the mass is conserved. This changes the angular momentum. The sum of angular momentum of all bodies changes/differences must be equal to the total difference in angular momentum between two concentric rings. The distribution function of dimension scales is equal to the distribution function of time scales, because the dimension scale divided by the local viscous velocity gives the time scale of a turbulent event. All events liberate energy in the form of a flickering flare. The distribution functions between pairs of rings are summed over the entire disc (from the inner disc radius $r_{\rm in}$ to the outer disc radius $r_{\rm out}$), which yields the final distribution function of flare time scales. Following this distribution function, flickering flares are randomly redistributed into a synthetic light curve of the same duration as the observations. Such synthetic light curves are subsequently analysed.

Finally, as mentioned above, nova-like systems should possess fully ionised accretion discs. Therefore, for basic disc parameters we used the standard \citet{shakura1973} model (see e.g. \citealt{frank1992}).

Examples of three light curves with MV\,Lyr system parameters from Section~\ref{mvlyr} with different disc parameters are depicted in Fig.~\ref{lc_examples}. The first two panels show light curves from an entire disc integrated (almost) from the white dwarf surface for two different outer disc radii. To study the multiple frequency behaviour of the observed light curves, we used our model to simulate also the longest viscous time scale variability. This is generated by turbulence at the outer disc rim, where the viscous velocity is lowest and turbulence is largest. This solution is based on the idea, that the outer disc accretion process is somehow dominant over the rest of the outer disc and adds additional variability to the light curve with its own characteristic statistics. For this purpose we simulated light curves with $r_{\rm out}$ equal to the disc radius and $r_{\rm in}$ slightly lower (bottom panel of Fig.~\ref{lc_examples}). Such radially very thin disc is imitating the outer disc rim. The final light curve is a sum of both models, i.e. "standard" full disc plus an outer disc rim\footnote{We do not have an exact physical model for the outer disc rim activity enhancement, hence we can not include it directly to the simulation procedure. Therefore, we simulated both situations separately and searched for the characteristic break frequency individually for every model.}. Therefore, we simulated two solutions with different parameters. A full disc constructed from the vicinity of the white dwarf (depending on the truncation) to the outer edge and an outer disc rim constructed around the outer disc radius with very small radial thickness. The radii of both models are summarised in Table.~\ref{results_tab}.

\section{Simulations and Analysis}
\label{simulations}

\begin{figure}
\includegraphics[width=77mm,angle=-90]{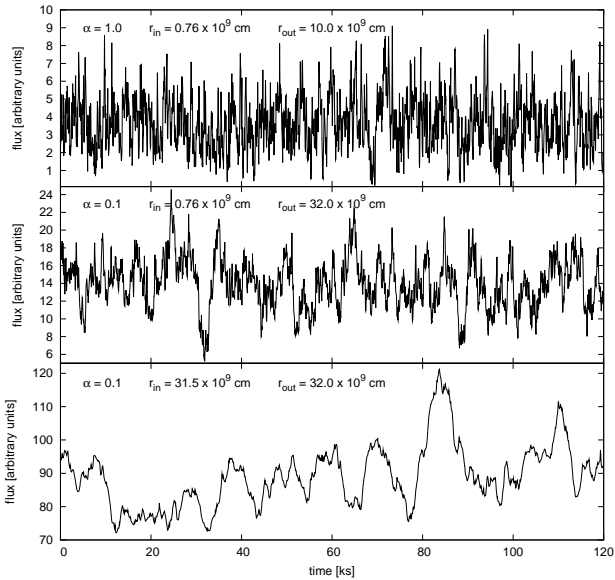}
\caption{Three examples (120\,ks sequence) of synthetic light curves assuming a mass accretion rate of $1.9 \times 10^{17}$\,g\,s$^{-1}$ for different values of the $\alpha$ parameter and disc radii. The number of flares per light curve is the same in all three cases.}
\label{lc_examples}
\end{figure}

As demonstrated in \citet{dobrotka2014} the number of flares per light curve does not matter, while the relative number of flares with different time scales is important. Therefore, we choose the lowest possible number of flares to minimise computer time. This limit is set by the "fastest" solution (top panel of Fig.~\ref{lc_examples}) with the shortest time scale of variability, because the sum of all flares must construct a continuous light curve without zero flux intervals\footnote{Shortest flares also occupy the shortest interval on the time axis. Therefore, the shorter each flare, the higher the number of flares needed to cover the entire time axis of the light curve. A low number of flares results in non covered zero flux intervals.}. The simulated light curve has the same duration (5.275\,d, 455760\,s) as observed light curve subsamples analysed by \citet{scaringi2012b} and we used approximately the same sampling as the \kepler~data (60\,s). 5 light curves were used per mean PDS, which we fitted with a Lorentzian function
\begin{equation}
P(\nu) = a_0 + \frac{a \Delta}{\pi} \frac{1}{\Delta^2 + (f - f_0)^2},
\end{equation}
where $a_0$ and $a$ are constants, $\Delta$ is the half-width at half maximum, $f$ is the frequency and $f_0$ is the studied break frequency. For PDS calculation we used Lomb-Scargle (\citealt{scargle1982}) algorithm and we binned the mean PDS into 0.1dex in the logarithmic frequency scale. We repeated this procedure 100 times and calculated a mean value of the frequency $f_0$.

Every PDS study is highly model dependent and mainly the power law slope can be difficult to model. The multicomponent PDS is a sum of more power laws which strongly affect the resulting shape. In \citet{dobrotka2014} the model yields a lower break frequency, but not the power law slope. This was explained by shorter time scale variability characterised by the second higher break frequency superposed on longer time scale variability (characterised by the lower break frequency). However, the break frequency describes the characteristic time scale which has the real physical meaning and is unchanged in the multicomponent PDS (unless blended with another adjacent too close break frequency). Therefore, we concentrate only on the break frequency of the PDS.

The next step is the calculation of the rms-flux relation. The absolute rms amplitude of variability is defined as square-root of the variance, i.e.
\begin{equation}
\sigma_{\rm rms} = \sqrt{\frac{1}{N - 1} \sum^N_{i = 1} (x_i - \overline{x})^2},
\end{equation}
where $N$ is the number of points, $x_i$ is the $i$-th point and $\overline{x}$ is the mean value of all $x_i$. We subdivided the 5 light curves used for the mean PDS calculation into small parts each containing 10 points (equivalent to 10 minutes as used in \citealt{scaringi2012a}) of the light curve. Subsequently for each small part we calculated the corresponding rms and mean flux. The binned data by flux were fitted with a linear function.

As input parameters we need $\alpha$, the white dwarf mass, the inner and outer disc radius and mass accretion rate. The $\alpha$ parameter in a hot ionised disc should be of the order of 0.1 (see e.g. \citealt{kato1998}, \citealt{lasota2001}). Therefore, we probed an interval from 0.1 to the most extreme value of 1.0. We estimated the white dwarf radius from the mass following \citet{nauenberg1972}, yielding $0.75 \times 10^9$\,cm. Therefore, an almost non-truncated accretion disc should have a comparable inner radius $r_{\rm in}$ (we adopted $0.76 \times 10^9$\,cm). As the outer disc radius $r_{\rm out}$ we used an order estimate of the typical value in cataclysmic variables of $10^{10}$\,cm (see e.g. \citealt{warner1995}), and two values depending on the fraction of the primary Roche lobe described in Section~\ref{mvlyr}.

All simulated models with combined input radii values are summarised in Table~\ref{results_tab} (models A$_i$, B$_i$ and C$_i$). Every model was calculated with three mass accretion rate values 1,2 and $3 \times 10^{17}$\,g\,s$^{-1}$. Finally we varied only the $\alpha$ parameter from 0.1 to 1.0.

\section{Results}
\label{results}

\begin{figure*}
\includegraphics[width=85mm,angle=-90]{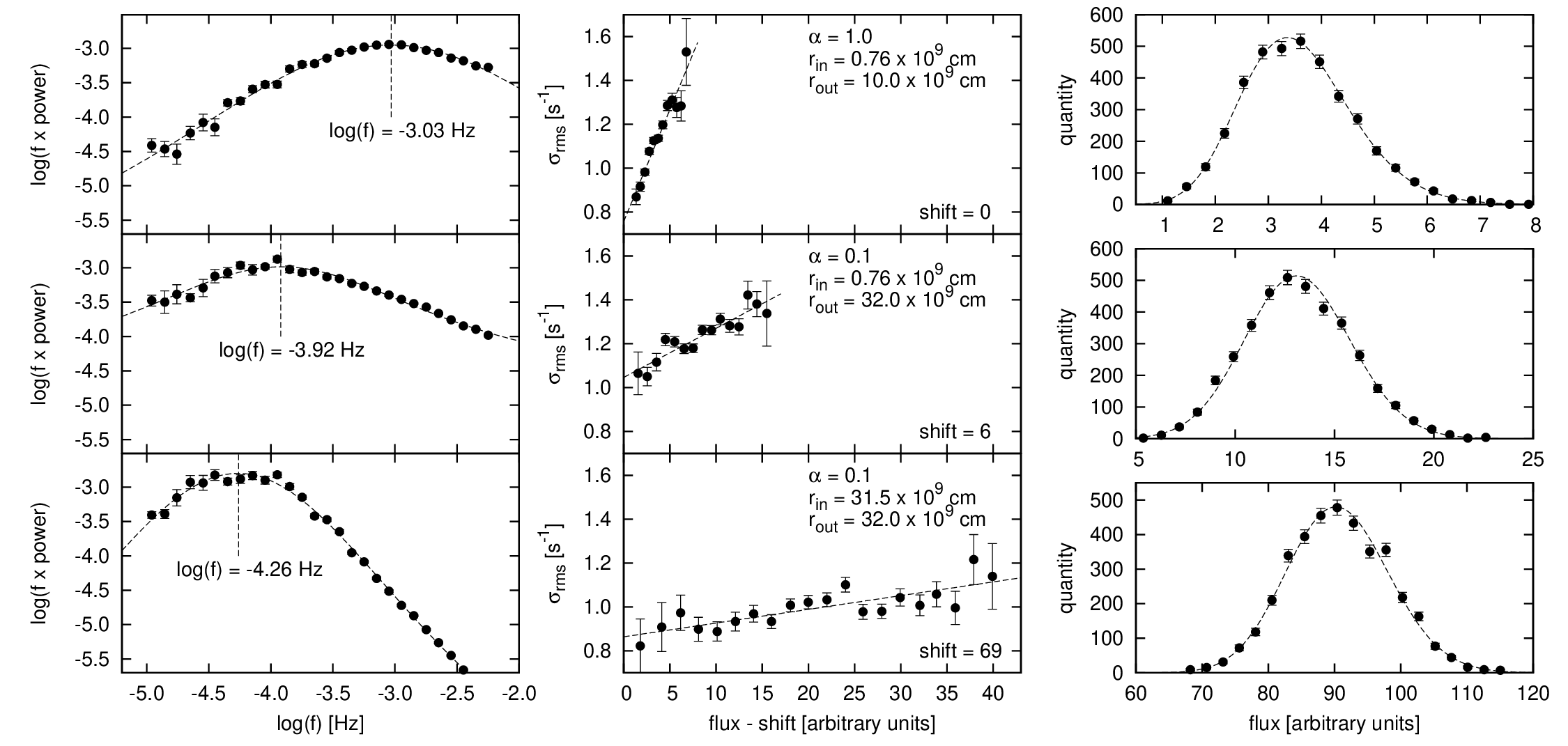}
\caption{Left column - three examples of mean PDS calculated from 5 simulated light curves (one light curve example for each case is shown in Fig.~\ref{lc_examples}) using different disc parameters. The dashed line is the Lorentzian fit. Middle column - corresponding rms-flux relations with linear fit (dashed line). The data with the fits are horizontally shifted in the middle and bottom case. Right column - corresponding binned flux distribution with a log-normal fit as dashed line.}
\label{rms-flux_examples}
\end{figure*}

Fig.~\ref{lc_examples} shows three examples of simulated light curves (120\,ks sequence of the full light curve). The variability time scale difference between all cases can clearly be seen. Also the relative flux is higher in light curves with longer time scale variability, because wider flares are summed, while an equal number of flares per light curve is used. Corresponding PDSs, rms-flux proportionality linear relations and flux distribution with Lorentzian, linear and log-normal fits, respectively, are depicted in Fig.~\ref{rms-flux_examples}. Clearly, smaller variability time scales yield steeper rms-flux relations and flux distribution more consistent with log-normality. We tried also Gaussian fits to the flux distribution for comparison, and except for the longest time scale variability case, the log-normal fits yield considerably better reduced $\chi^2$: 0.79 vs. 5.49 and 1.81 vs. 2.50 for the shortest and medium time scales, respectively. The longest time scale case yields very similar fits with both reduced $\chi^2$ of 1.7. Any further detailed comparison with observed values and subsequent quantitative study is not possible, because our simulation method does not work with real flux units.

\begin{figure}
\includegraphics[width=74mm,angle=-90]{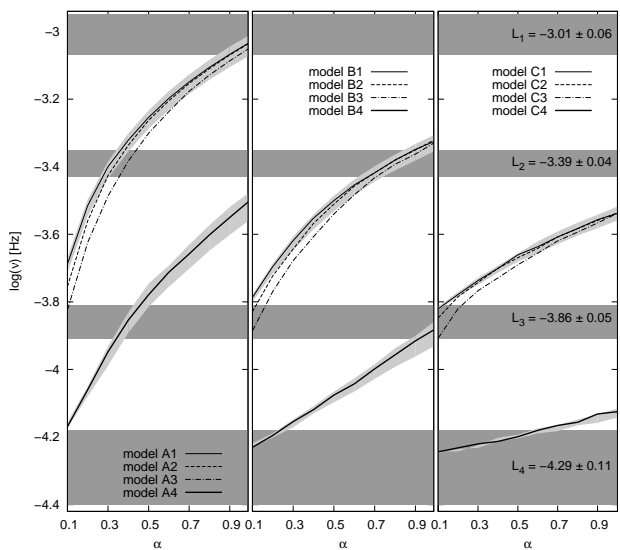}
\caption{Simulated break frequencies for different disc radii (mean values after 100 simulations). Thin lines are full disc simulations and thick lines are the outer disc rim simulations. Lighter shaded areas are intervals of simulated break frequencies taking into account a mass accretion rate interval from 1 to $3 \times 10^{17}$\,g\,s$^{-1}$. The lines indicate mass accretion rates of $1.9 \times 10^{17}$\,g\,s$^{-1}$. Darker shaded areas are error intervals of observed values deduced in Section~\ref{mvlyr}.}
\label{results_fb}
\end{figure}

The simulated mean values of break frequencies are depicted in Fig.~\ref{results_fb}. The errors of the mean are of the order of 0.01, hence negligibly small and not displayed for clarity of the figure. The lines are results with mass accretion rate of $1.9 \times 10^{17}$\,g\,s$^{-1}$, but the lighter shaded areas show the simulated mean break frequency intervals using different mass accretion rate values\footnote{Not shown for all models for clarity of the figure.} from 1 to $3 \times 10^{17}$\,g\,s$^{-1}$. Apparently, the influence is relatively weak, mainly in the case of low frequency variations. The darker shaded areas are the error intervals of the observed break frequencies derived in Section~\ref{mvlyr}.

In the upper half of Table~\ref{results_tab} we list the $\alpha$ intervals, where simulated values (only with the mass accretion rate of $1.9 \times 10^{17}$\,g\,s$^{-1}$) match the observed intervals. The bottom part of Table~\ref{results_tab} lists the combinations, where $\alpha$ values satisfying the observations agree from both full disc and outer disc rim simulations, because we assumed the $\alpha$ parameter to be constant in the full disc. Therefore, only those models where $\alpha$ values overlap from both simulations make sense. Finally, radial variations of $\alpha$ between the disc and the outer disc rim are worth studying, but this is beyond the scope of this paper (see Section~\ref{constant_alpha} for discussion).
\begin{table*}
\caption{First half - List of $\alpha$ values where the simulated mean break frequency agrees with the observed interval. Second half - $\alpha$ values (from top part) where both models (full disc and outer disc rim) yield the same (overlapping) intervals.}
\begin{center}
\begin{tabular}{lllll}
\hline
\hline
$r_{\rm out}$ & \multicolumn{4}{c}{$r_{\rm in}$}\\
\noalign{\smallskip}
\cline{2-5}
\noalign{\smallskip}
($10^9$\,cm) & $0.76 \times 10^9$\,cm & $1.50 \times 10^9$\,cm & $2.00 \times 10^9$\,cm & $r_{\rm out} - 0.5 \times 10^9$\,cm\\
& full disc & full disc & full disc & outer disc rim\\
\hline
10 & model A1 & model A2 & model A3 & model A4\\
& $L_1$: 0.89 - 1.00 & $L_1$: 0.89 - 1.00 & $L_1$: 0.95 - 1.00 & $L_1$: ---\\
& $L_2$: 0.28 - 0.36 & $L_2$: 0.30 - 0.38 & $L_2$: 0.35 - 0.44 & $L_2$: ---\\
& $L_3$: --- & $L_3$: --- & $L_3$: 0.10 - 0.11 & $L_3$: 0.34 - 0.46\\
& $L_4$: --- & $L_4$: --- & $L_4$: ---  & $L_4$: ---\\
\hline
18 & model B1 & model B2 & model B3 & model B4\\
& $L_1$: --- & $L_1$: --- & $L_1$: --- & $L_1$: ---\\
& $L_2$: 0.68 - 0.91 & $L_2$: 0.68 - 0.91 & $L_2$: 0.71 - 0.93 & $L_2$: ---\\
& $L_3$: --- & $L_3$: 0.10 - 0.12 & $L_3$: 0.10 - 0.16 & $L_3$: 0.93 - 1.00\\
& $L_4$: --- & $L_4$: --- & $L_4$: --- & $L_4$: 0.10 - 0.24\\
\hline
32 & model C1 & model C2 & model C3 & model C4\\
& $L_1$: --- & $L_1$: --- & $L_1$: --- & $L_1$: ---\\
& $L_2$: --- & $L_2$: --- & $L_2$: --- & $L_2$: ---\\
& $L_3$: 0.10 - 0.12  & $L_3$: 0.10 - 0.16 & $L_3$: 0.12 - 0.22 & $L_3$: ---\\
& $L_4$: --- & $L_4$: --- & $L_4$: --- & $L_4$: 0.10 - 0.60\\
\hline
\hline
10 & model A1 & model A2 & model A3 & model A4\\
& $L_2$: 0.28 - 0.36 & $L_2$: 0.30 - 0.38 & $L_2$: 0.35 - 0.44 & ---\\
& $L_3$: 0.34 - 0.46  & $L_3$: 0.34 - 0.46 & $L_3$: 0.34 - 0.46 & ---\\
\hline
18 & model B1 & model B2 & model B3 & model B4\\
& $L_3$: --- & $L_3$: 0.10 - 0.12 & $L_3$: 0.10 - 0.16 & ---\\
& $L_4$: ---  & $L_4$: 0.10 - 0.24 & $L_4$: 0.10 - 0.24 & ---\\
\hline
18 & model B1 & model B2 & model B3 & model B4\\
& --- & --- & $L_2$: 0.71 - 0.93 & ---\\
& --- & --- & $L_3$: 0.93 - 1.00 & ---\\
\hline
32 & model C1 & model C2 & model C3 & model C4\\
& $L_3$: 0.10 - 0.12 & $L_3$: 0.10 - 0.16 & $L_3$: 0.12 - 0.22 & ---\\
& $L_4$: 0.10 - 0.60  & $L_4$: 0.10 - 0.60 & $L_4$: 0.10 - 0.60 & ---\\
\hline
\end{tabular}
\end{center}
\label{results_tab}
\end{table*}

\section{Discussion}
\label{discussion}

In this work we simulated synthetic light curves of the nova-like system MV\,Lyr in order to study fast optical variability detected in \kepler~data. Owing to the complicated structure of the observed PDS, which is probably a superposition of 4 different PDS components, we focused our study only on the break frequencies. The goal was to compare the simulated and observed values for various system parameters. We simulated fast variability generated by the full disc and by a radially thin outer disc rim. We used the latter to search for the longest time scale variability generated by viscous processes.

In general smaller outer disc radii yield shorter simulated time scales of variability, while larger inner disc radii yield longer simulated time scales. However, the simulated break frequencies are only weakly sensitive to changes of $r_{\rm in}$. This break frequency behaviour is different from what is found for similar simulations of the power law slope in the case of X-ray binaries \citet{dobrotka2014b}, where the results are highly sensitive to the inner disc radius, and are insensitive to changes of the outer disc radius. Although in \citet{dobrotka2010} and \citet{dobrotka2012} it is shown that the break frequency and power law index are correlated in our simulations, both PDS parameters have different sensitivity to the disc radius.

We investigated models with three inner and three outer disc radii. Despite the sensitivity of our simulation to the outer disc radii values, the qualitative difference between our simulations with $r_{\rm out} = 1.8$ and $3.2 \times 10^{10}$\,cm are weak, and only extremely small $r_{\rm out} = 1 \times 10^{10}$\,cm completely change the interpretation. In the following, we discuss all simulated models in detail.

\subsection{Small disc model}
\label{small_disc}

Small disc models (A1-A4 in Table~\ref{results_tab}, first half) with $r_{\rm out} = 1 \times 10^{10}$\,cm ($0.3 \times r_{\rm PRL}$) are able to explain the observed break frequencies $L_1$, $L_2$ or $L_3$. $L_1$ is generated only by the full disc model (the outer disc rim does not fit any observed break frequency) and yields extreme $\alpha$ values between 0.9 and 1.00. This is suspiciously large because the typically observed values in hot discs lie between 0.1 and 0.4 (\citealt{king2007}). However, this result agrees with an $\alpha$ parameter deduced for the recurrent nova and symbiotic system T\,CrB (\citealt{dobrotka2010}) and for the studied MV\,Lyr system from previous PDS modelling (\citealt{scaringi2014}). But the latter has a different disc interpretation, where the disc is believed to be geometrically thick (we discuss this interpretation later).

The more plausible model which agrees with typically observed values of $\alpha$ explains the observed break frequencies $L_2$ and $L_3$ (A1-A3 in Table~\ref{results_tab}, first half), where the latter is generated by the outer disc rim. In this model the observed break frequencies $L_1$ and $L_4$ remain unexplained.

Furthermore, assuming an inner disc truncation up to $2 \times 10^9$\,cm, the full disc model (A3 in Table~\ref{results_tab}, first half) is able to reproduce the observed frequency $L_3$ for more reasonable values of $\alpha$, but with the outer disc rim not satisfying (for the same $\alpha$) any of the observed break frequencies. Such solution is also untrustworthy due to large values of the inner disc, which we discuss later.

Assuming a disc with constant $\alpha$, the resulting simulated frequencies are $L_2$ and $L_3$ with $\alpha$ values of 0.3 - 0.5 for all disc models (A1-A3 in Table~\ref{results_tab}, second half).

Finally, a small disc with an outer disc radius equal to approximately $0.29 \times r_{\rm PRL}$ seems to be too small and such a solution needs strong observational support. Therefore, we do not expect this solution to work.

\subsection{Large disc models}

More realistic disc diameters between 0.5 and $0.9 \times r_{\rm PRL}$ yield almost the same results (models B1-B4, C1-C4 in Table~\ref{results_tab}, first half), with weak dependence on the inner disc radius. The fitted observed frequencies are $L_3$ and $L_4$ for full disc models (B1-B3 and C1-C3 in Table~\ref{results_tab}, first half) and $L_3$ and $L_4$ for outer disc rim (models B4 and C4 in Table~\ref{results_tab}, first half). The resulting $\alpha$ values are again cumulated in two intervals, i.e. small values of 0.1 - 0.2 and larger values of 0.7 - 1.00.

Assuming again a disc with constant $\alpha$ yields simulated frequencies $L_3$ and $L_4$ for $r_{\rm out} = 18 \times 10^{10}$\,cm (B2 and B3 in Table~\ref{results_tab}, second half) with again two $\alpha$ intervals of 0.1 - 0.2 and 0.7 - 1.00, and for $r_{\rm out} = 32 \times 10^{10}$\,cm (C1 and C3 in Table~\ref{results_tab}, second half) with $\alpha$ of about 0.1 - 0.6.

Finally, all models with the lower $\alpha$ intervals are more attractive because they are observationally supported as summarised by \citet{king2007}.

\subsection{Inner disc truncation}

In our simulations we used three inner disc radii (A1-A3, B1-B3 and C1-C3 in Table~\ref{results_tab}) from the very close white dwarf vicinity (non truncated disc) to $2 \times 10^9$\,cm (truncated disc). The sensitivity of our simulation to this disc parameter is weak, and with four possible break frequencies we can not deduce whether or not the disc is truncated. \citet{scaringi2014} analysed the same data, and specifically modelled the $L_1$ component of this work. The authors concluded that the disc is an optically thin hot geometrically thick corona from the white dwarf to an outer radius of $8 \times 10^9$\,cm with a potential geometrically thin disc inside, i.e. a sandwich model.\\

Such corona in accreting systems appears in the central disc, when the mass accretion rate drops to low values where the cooling becomes inefficient and the temperature rises strongly causing evaporation of disc material (see e.g. \citealt{meyer1994}, \citealt{mineshige1998}, \citealt{meyer2007}). Once the material is evaporating, the geometrically thin disc disappears and the corona forms instead (see Fig.~1 of \citealt{meyer2007}). Therefore, the outer radius of the geometrically thick corona must be similar to the inner radius of the geometrically thin disc. In nova-like systems, the geometrically thin disc is believed to be in the hot ionised stage except during steep declines (antinova phenomenon, see \citealt{warner1995}) and the inner disc radius should be considerably smaller than in dwarf novae during quiescence (if truncated, see later in this paragraph). \citet{schreiber2003} modelled the outburst cycle for the dwarf nova SS\,Cyg and found a successful solution with an inner disc truncation of $2 \times 10^9$\,cm during quiescence. For the nova-like system KR\,Aur, \cite{dobrotka2012} found an inner disc radius in the interval of $0.88 - 1.67 \times 10^9$\,cm, which is considerably smaller than in SS\,Cyg in quiescence as expected. Therefore, an inner disc radius up to $2 \times 10^9$\,cm is rather too large for a system with a fully ionised disc. Furthermore, a truncated disc with $r_{\rm in} = 2.5 \times 10^9$\,cm from \citealt{catalan1995} is suggested by eclipse fitting in HT\,Cas during the rise to outburst (\citealt{ioannou1999}), but it should be lower during maximum/outburst. In the study of V2051\,Oph (\citealt{baptista2004}) and V4140\,Sgr (\citealt{borges2005}) the inner disc hole can be estimated from the innermost point in the radial brightness temperature distribution estimates deduced from eclipse mapping. In V2015\,Oph, the disc seems not to be truncated while in V4140\,Sgr the inner disc radius is roughly estimated of about $1.4 \times 10^9$\,cm. Clearly, not only the inner disc radius, but the disc truncation in general remains unclear in some systems. Nevertheless, a disc truncation of $8 \times 10^9$\,cm in nova-like systems with a hot fully-ionised disc is rather problematic, because it gives considerably larger values than in quiescent systems. The sandwiched model proposed by \citet{scaringi2014} is therefore a valid alternative solution, i.e. the hot corona is not identical with the thin disc inner hole, but the thin disc can be fully developed down to the white dwarf surface and a corona is formed above the disc material from the white dwarf to an outer radius of $8 \times 10^9$\,cm.

\subsection{Final model} 

The model proposed by \citet{scaringi2014} is only one alternative. Fragmentation of the accretion flow in the boundary layer between the inner disc and the stellar surface can be a different possible scenario (see \citealt{frank1992} for review). A similar interpretation has also been discussed for RU\,Peg during quiescence (\citealt{dobrotka2014}). This interface region is subject to a strong tangential velocity gradient, and it therefore satisfies the basic conditions for Kelvin-Helmholtz instability. Such fragmenting/unstable strongly turbulent flow can generate fast X-ray variability, which can subsequently be reprocessed by the geometrically thin disc.

There is only one exact method to test the proposed reprocessing hypothesis, i.e., direct X-ray observations. In case of $L_1$ generated by reprocessing, this break frequency should be detected in X-rays. Furthermore, with simultaneous optical and X-ray observations a time-lag for the shortest variability should be detected. A similar lags have been found by \citet{scaringi2013} in optical red and blue bands. They detected red/negative lags where the redder bands lag behind the bluer ones which supports reprocessing. Therefore, a long enough X-ray observation of MV\,Lyr would be needed \footnote{Ideas about such observation characteristics as duration, analysis procedure etc. can be taken from \citet{dobrotka2014}}. However, a positive detection of reprocessing still leaves an ambiguity about the origin of the frequency $L_1$ as predicted either by the sandwiched model with a geometrically thick disc or the boundary layer model.

Furthermore, it is worth emphasising that our finding of $L_1$ in small disc models (discussed in Section~\ref{small_disc}) is consistent with the findings by \citet{scaringi2014} despite the different disc interpretation (geometrically thin vs. thick disc). Both the outer disc radii and $\alpha$ values are very similar. This encourages us to adapt our method to geometrically thick discs and search for the $L_1$ frequency and find potential support for the sandwiched model of \citet{scaringi2014}.

Finally, implementing the corona interpretation by \citet{scaringi2014} in the overall model and assuming the $\alpha$ parameter to be between 0.1 and 0.4 as typically deduced from observations (\citealt{king2007}), the interpretation of the observed variability in MV\,Lyr would be as follows; the frequency $L_1$ is generated by a fragmented accretion flow in the X-ray corona or boundary layer (generating X-ray variability reprocessed by the disc), and $L_3$ with $L_4$ are generated by a turbulent accretion flow in the full disc and outer disc rim, respectively. However, the frequency $L_2$ remains unexplained.

\subsection{Different $\alpha$ parameters in the disc and outer disc rim}
\label{constant_alpha}

It is believed that turbulence in an accretion disc is produced by the magnetorotational instability (MRI, \citealt{balbus1998}). Therefore, the driving mechanism of the studied turbulent flow is the magnetic field. In active discs, hydrogen is fully ionised and the magnetic field is coupling effectively to the gas and a saturation limit of the MRI can yield a constant $\alpha$ parameter. This is the reason why many studies assume a constant value of $\alpha$ during the hot stage as we did in our work (Table~\ref{results_tab}, second half). However, this value in the outer rim could be different from those in the disk, since the way of global coupling is not the same at the outer rim and within the disk, since there is no further coupling (at the rim) with material outside the outer rim. This is not conclusive and merely a speculation, but more detailed investigation is beyond the scope of this paper. But this suggests, that models with different values of $\alpha$ can not be definitely ruled out. Qualitatively, this statement does not change the conclusion of our work. At least it can be a suggestion, why the outer disc region activity is enhanced and contributes much more to the variability statistics than in the "standard" disc model.

\subsection{Rms-flux relation}

The typical rms-flux proportionality linear relation with log-normal flux distribution detected in many accreting binary systems is a characteristic of our model as well. In general, the longer the time scale variability, the lower the steepness of the linear dependence. Finally, the accretion flow fragmentation into turbulent bodies transporting the angular momentum within the disc generates mass accretion fluctuations which propagate further through the disc. Therefore, this can confirm the general knowledge (\citealt{uttley2005}, \citealt{ingram2013}), that the typical rms-flux relation is mainly generated by the mass accretion rate fluctuations propagating in the accretion disc.

Any detailed comparison of the simulated linear steepness with the observed data is not possible, because our method only simulates the flux in arbitrary units. As already mentioned, it is only a statistical model.

In general higher rms are generated by higher mass accretion rates. During initial testing of our method, we obtained higher rms (of small light curve segments or of the whole light curve) when using a larger number of simulated flares. This is compatible with a higher mass accretion rate, because larger amounts of angular momentum need to be transported, which results in a larger number of flares. In fact, in our simulations we keep the mass accretion rate as input parameter constant, but the flares are redistributed randomly, which naturally generates fluctuations in the concentration of flares per light curve segment. The assumed mass accretion rate is a mean value per light curve. Therefore, a larger rms in a light curve segment means that a larger number of flares is present in the segment, which corresponds to larger mass accretion rate expressed by larger flux. The simulated rms-flux relation has a physical explanation and is not an artifact of our method.

Finally, the rms-flux relation is not only a linear correlation, but it is a proportionality rms $\propto$ flux, i.e. zero flux means no flares in the constructed synthetic light curve and this obviously yields zero rms. Therefore, any deviation from this trend seen in Fig.~\ref{rms-flux_examples} is an intrinsic numerical artefact of our simulation procedure which can be improved in the future. Furthermore, the gradient of the rms-flux relation might depend on the timescale onto which the rms-flux relation is calculated from, but detailed MV\,Lyr data analysis is out of scope of this paper.

\section{Summary}
\label{summary}

We modelled fast optical variability using parameters of the nova-like system MV\,Lyr. The results can be summarised as follows:

(i) The fragmented turbulent accretion flow generates the typical linear rms $\propto$ flux relation. The steepness of this linear dependence is higher for shorter time scale variability.

(ii) In general larger rms are generated by a larger number of superposed flares. This corresponds to higher mass accretion rates expressed as larger flux. Therefore, variable rms agrees with mass accretion rate fluctuations.

(iii) Assuming typical disc radii of cataclysmic variables and related objects between 0.5 and 0.9 times the primary Roche lobe and taking the typical observed $\alpha$ parameter to be of 0.1 - 0.4 (\citealt{king2007}), our simulations of variability coming from the entire accretion disc are able to explain the observed break frequency $L_3$ (\citealt{scaringi2014}), while the simulations of variability generated by an outer disc rim explains the observed frequency $L_4$.

(iv) The previous point suggests that viscous processes are somewhat enhanced at the outer disc rim.

(v) Using previous work of \citet{scaringi2012b}, the break frequency $L_1$ can be explained by a geometrically thick corona radiating in X-rays, where optical variability is generated by reprocessing of the X-rays by the disc. Despite the very different disc geometry used by us (geometrically thin disc), we find consistent parameters also with the \citet{scaringi2014} model. In such case only the observed frequency $L_2$ remains unresolved.

\section*{Acknowledgements}

AD was supported by the Slovak Academy of Sciences Grant No. 1/0511/13. SM was supported in part by the Grant-in-Aid of Ministry of Education, Culture, Sports, Science, and Technology (MEXT) (22340045, SM) and by the Grant-in-Aid for the global COE programs on The Next Generation of Physics, Spun from Diversity and Emergence from MEXT. We are grateful to the HPC centre at the Slovak University of Technology in Bratislava, which is a part of the Slovak Infrastructure of High Performance Computing (SIVVP project, ITMS code 26230120002, funded by the European region development funds, ERDF), for the computational time and resources made available.

\bibliographystyle{mn2e}
\bibliography{mybib}

\label{lastpage}

\end{document}